\documentclass[12pt, prd, showpacs]{revtex4}
\usepackage{amssymb}
\usepackage{amsmath}

\input{tcilatex}

\begin{document}

\title{High energy particle collisions near the bifurcation surface}
\author{Oleg B. Zaslavskii}
\affiliation{Kharkov V.N. Karazin National University, 4 Svoboda Square, Kharkov, 61077,
Ukraine}
\email{zaslav@ukr.net}

\begin{abstract}
We consider generic nonextremal stationary dirty black holes. It is shown
that in the vicinity of any bifurcation surface the energy of collision of
two particles in the centre of mass frame can grow unbound. This is a
generic property that, in particular, includes collisions near the inner
black hole horizon analyzed earlier by different methods. The similar
results are also valid for cosmological horizons. The case of the de Sitter
metric is discussed.
\end{abstract}

\maketitle

% It is always \today, today, but any date may be explicitly specified

%\newpage

\section{Introduction}

The effect of acceleration of particles by black holes (called the BSW
effect according to the names of its authors) was discovered quite recently 
\cite{ban} and is now under active study. Apart from significant pure
theoretical interest, its potential astrophysical applications also attract
attention (see, e.g., recent papers \cite{insp}). The essence of this effect
consists in the unbound growth of the energy $E_{c.m.}$ in the centre of
mass frame of two colliding particles. Mainly, it concerns the collision
near the event horizon. Meanwhile, this effect was discussed also in a quite
different situation - near the inner black hole horizon. It turned out \cite%
{in} that a crucial role is played in that case by the bifurcation
two-dimensional surface where the future and past horizons meet. Namely, for
the BSW effect to occur, particles should collide in the vicinity of this
surface.

The goal of the present work is to pay attention that \textit{any }%
bifurcation surface can lead to the BSW effect. In particular, it includes
the case of the inner black hole horizon discussed previously \cite{in} - 
\cite{gpgc}. But this is also true even if there is no an inner horizon (for
example, in the Schwarzschild case). It is worth reminding that the BSW
effect was found for extremal horizons \cite{ban} (which do not have a
bifurcation sphere at all). Later on, it was shown that it occurs also near
the nonextremal event horizons \cite{gpjl}, \cite{prd} but not on the
bifurcation surface. (Hereafter, when mentioning the properties of the BSW
effect near the horizon, we imply just one of these two cases, if the
bifurcation surface is not mentioned explicitly.) The similar effect was
discussed also near singularities \cite{nk}. Now, we add a new object to
this list of possible particle accelerators in a strong gravitational field.

Some reservations are in order. One of examples of the bifurcation point
occurs inside the event horizon where two inner ones intersect. It is known
that near such horizon a strong instability develops \cite{pi} (see also,
e.g. the reviews \cite{ori}, \cite{ham}), so one may ask if it makes sense
to study the processes near the object which cannot probably exist in a real
world. In out view, there are several points to motivate studies of the BSW
effect near the bifurcation surface. (i) It may happen that the BSW effect
itself can contribute to further instability of inner horizons but this kind
of instability is quite different from that found in \cite{pi}. If so, this
can be considered as an additional argument against the existence of such
horizons. It would be of interest to evaluate the relative role of both
effects that seems to be a separate issue. But the first step here is to
show that the corresponding property (BSW effect) is indeed inherent to any
bifurcation point. The similar conclusions concern also white holes (see
Sec. 15.2 of \cite{fn}). (ii) The area of potential applications is not
restricted by black hole inner horizons. As follows from consideration
below, our results apply not only to the inner black hole horizons but also
to cosmological ones (the de Sitter spacetime is one example). Such objects
not only exist but play an obviously crucial role in modern cosmology. (iii)
Moreover, our approach and corresponding results apply to any space-time
with nontrivial causal structure where different nonextremal horizons
alternate. (iv) Independently of direct applications, the BSW process is a
nontrivial gravitational effect.\ The list of objects to which it applies is
gradually growing (see. e.g. the recent work \cite{rec}), so it is important
problem to understand in which situations and under which conditions it
happens. In our view, it looks reasonable to separate two issues - the
nature of the BSW effect as such and its potential relevance in realistic
astrophysics. The present work concerns the first issue only.

The results of the present work apply to quite generic "dirty" horizons. In
astrophysical circumstance a real black hole is not in vacuum but is
surrounded by matter. Therefore, it does not have a simple, say, Kerr or
Reissner-Nordstr\"{o}m form. Although exact solutions for such systems are
absent, universality of black hole physics enabled to derive some general
relationships. Interest to such objects is also connected with such general
issues as the black hole entropy, its relation to the horizon symmetries,
etc. (see \cite{vis1}, \cite{vis2}, \cite{tan}, \cite{curly}).

\section{Two versions of BSW effect}

The crucial role is played in the BSW effect by division of all possible
trajectories of particles into two classes - the so-called "critical"
particles (that implies special relationship between particle's parameters)
and so-called "usual" particles - all other ones \cite{in}, \cite{prd}, \cite%
{cqg}. More precise definitions will be given below. As is shown in \cite{in}%
, it is impossible to have configurations that satisfy simultaneously the
following properties: (i) $E_{c.m.}$ is infinite ("the strong version of the
BSW effect"), (ii) collision does occur in some point. In particular,
property (i) requires particle 1 (critical) to pass through the bifurcation
point while particle 2 is to be usual and this excludes condition (ii). In
this sense, the "kinematic censorship" forbids actual infinity in any
physical event. However, if collision takes place not exactly on the
bifurcation surface but somewhere in its vicinity, $E_{c.m.}$ turns out to
be finite but can be made as large as one likes. This was called in \cite{in}
the "weak version" of the BSW effect. In what follows, we imply just this
version without further reservations. It is the proof that the bifurcation
surface can serve as a particle accelerator that we now turn to.

\section{BSW effect near bifurcation point}

This proof is surprisingly simple. We follow the general analysis suggested
in \cite{cqg} but apply it not to the event horizon but, instead, to the
bifurcation surface. To make presentation self-closed, I remind briefly main
ingredients of the approach. Let colliding particles have the masses $m_{1}$%
, $m_{2}$. Then, by definition,%
\begin{equation}
E_{c.m.}^{2}=-P_{\mu }P^{\mu }\text{, }P^{\mu }=m_{1}u_{(1)}^{\mu
}+m_{2}u_{(2)}^{\mu }
\end{equation}%
is the total momentum. Hereafter, subscript (i=1,2) labels the particles.
Then,%
\begin{equation}
E_{c.m.}^{2}=m_{1}^{2}+m_{2}^{2}+2m_{1}m_{2}\gamma  \label{e}
\end{equation}%
where%
\begin{equation}
\gamma =-u_{(1)}^{\mu }u_{\mu (2)}  \label{gamma}
\end{equation}%
has the meaning of the Lorentz gamma factor of particles' relative motion.
The BSW effect exists if $\gamma $ can be made arbitrarily large.

Using the basis consisting of null vectors $l^{\mu }$ and $N^{\mu }$ and the
space-like vectors $a^{\mu }$ and $b^{\mu }$ orthogonal to them, one can
write%
\begin{equation}
u_{(i)}^{\mu }=\beta _{i}N^{\mu }+\frac{l^{\mu }}{2\alpha _{i}}+s_{(i)}^{\mu
}  \label{u}
\end{equation}%
where $s^{\mu }=s_{a}a^{\mu }+s_{b}b^{\mu }$ and, for definiteness, we use
normalization $l^{\mu }N_{\mu }=-1$. Then,%
\begin{equation}
\beta _{i}=-u_{(i)}^{\mu }l_{\mu }\text{, }2\alpha _{i}=-[u_{(i)}^{\mu
}N_{\mu }]^{-1}\text{.}  \label{ba}
\end{equation}%
The normalization condition $u_{(i)}^{\mu }u_{\mu (i)}=-1$ gives us%
\begin{equation}
\frac{\beta _{i}}{\alpha _{i}}=1+s_{(i)}^{\mu }s_{\mu (i)}\text{.}
\label{uu}
\end{equation}%
It follows from the above formulas that%
\begin{equation}
\gamma =\frac{1}{2}(\frac{\beta _{1}}{\alpha _{2}}+\frac{\beta _{2}}{\alpha
_{1}})-s_{(1)}^{\mu }s_{\mu (2)}\text{.}  \label{g}
\end{equation}%
Here, $s_{(i)}^{\mu }s_{\mu (i)}=s_{a(i)}^{2}+s_{b(i)}^{2}$, $s_{(1)}^{\mu
}s_{\mu (2)}=s_{a(1)}s_{a(2)}+s_{b(1)}s_{b(2)}$.

One can rotate the set of basis vectors ($l^{\mu }$, $N^{\mu }$, $a^{\mu }$, 
$b^{\mu }$) to ($\tilde{l}^{\mu }$, $\tilde{N}^{\mu }$, $\tilde{a}^{\mu }$, $%
\tilde{b}^{\mu }$). One can check that $\gamma (\alpha _{i}$, $\beta _{i}$, $%
s_{a(i)}$)=$\gamma (\tilde{\alpha}_{i},\tilde{\beta}_{i},\tilde{s}_{a(i)})$.
There is no need to perform straightforward calculations for this purpose,
it follows immediately from the scalar nature of $\gamma $ (\ref{gamma}) and
(pseudo)orthogonal character of both tetrads. Therefore, the choice of the
basis vector is a matter of convenience. In what follows, we assume that $%
s_{a}$ and $s_{b}$ are finite.

Then, if, say, $\alpha _{1}\rightarrow 0$ but $\alpha _{2}\neq 0$, $\gamma
\rightarrow \infty $. Particle 1 is called critical and particle 2 is called
usual. As for spacelike vectors $s_{(i)}^{\mu }s_{\mu (i)}\,\geq 0$, it is
seen from (\ref{uu}) that the condition $\beta _{i}=0$ entails $\alpha
_{i}=0 $ that will be used below. (It is worth noting that the reverse is
true with the additional condition that $s_{(i)}^{2}\equiv s_{(i)}^{\mu
}s_{\mu (i)}$ is finite. Then, the aforementioned conditions on $\alpha $
and $\beta $ become equivalent.)

In what follows, we assume that when the point of collision approaches the
future horizon, the light-like vector $l_{\mu }$ approaches its generator.
Let both particles move from infinity (or any location outside) towards the
future event horizon. Then, the condition $\beta _{1}=0$ (hence, also $%
\alpha _{1}=0$) gives rise to a special relationship between the particle's
parameters and in this sense selects a special type of trajectories among
all possible ones. Thus for the Reissner-Nordstr\"{o}m black hole it gives $%
E_{1}=q_{1}\varphi _{H}$ where $q$ is the electric charge of the particle, $%
\varphi _{H}$ is the electric potential, subscript "H" refers to the
horizon. For rotating black holes $E_{1}=\omega _{H}L_{1}$ where $L_{1}$ is
the particle's angular momentum, $\omega $ is the metric coefficient
responsible for rotation of the space-time (see \cite{cqg} for details).
Geometrically, the condition $\beta \rightarrow 0$ means that the component
of the velocity across the horizon vanishes and the critical particle cannot
cross it all. It only approaches it asymptotically, and an infinite proper
time is required to reach the extremal horizon \cite{ted}, \cite{gpjl}. (If
the horizon is nonextremal, the critical particle cannot reach it and the
explanation of the BSW effect somewhat changes \cite{gpjl}, \cite{prd}.)

Let us now consider the bifurcation surface instead of the future horizon.
First, we assume that a black hole is axially-symmetric and stationary. In
general, it is surrounded by matter (so it is "dirty") and does not
necessarily coincides with the Kerr metric or any other exact solution of
the field equations. Then, one can take advantage of the known properties
(see, for example, Sec. 5.1.10, 5.3.11 and 5.4.2 of textbook \cite{erik}).
On the horizon, $l^{\mu }=\xi ^{\mu }+\omega _{H}\eta ^{\mu }$ where $\xi
^{\mu }$ is the Killing vector $\xi ^{\mu }$ generates time translation and
the Killing vector $\eta ^{\mu }$ generates rotations. For a static metric, $%
\omega =0$. On the bifurcation surface, $l^{\mu }\rightarrow 0$. According
to (\ref{ba}), it means that $\beta _{1}\rightarrow 0$ and, from (\ref{uu}), 
$\alpha _{1}\rightarrow 0$ as well. Meanwhile, particle 2 is by assumption
usual, it foes not pass through the bifurcation surface, so $\alpha _{2}\neq
0$. As a result, (\ref{g}) grows unbound. This completes the proof. We
stress that in contrast to previous works \cite{in} - \cite{gpgc}, we did
not analyze the motion of test particles in the vicinity of the bifurcation
surface at all.

We can also do without the referring to the Killing vector. If, say, the
horizon lies at $r=r_{+}$, the normal vector $l_{\mu }\sim -\frac{\partial r%
}{\partial x^{\mu }}$ satisfies the condition $l_{\mu }l^{\mu }=0$. In the
coordinate system like

\begin{equation}
ds^{2}=-FdUdV+r^{2}d\omega ^{2}\text{,}  \label{f}
\end{equation}%
it reduces to $l_{U}l_{V}=0$. The bifurcation point, by its very meaning is
singled out by the condition that both factors tends to zero, so the vector $%
l_{\mu }$ itself vanishes (see below the explicit example for the
Reisnner-Nordstr\"{o}m metric). Generalization to spacetimes without
spherical symmetry is straightforward.

Both situations (the BSW effect near the event horizon and near the
bifurcation surface) are complimentary to each other in the following sense.
The key property of a critical particle $\alpha _{1}=\beta _{1}=0$ can be
realized in two ways. If one uses the Kruskal-like coordinates $U$ and $V$
in which the metric is nondegenerate, the property under discussion reduces
to the condition of the type $u^{U}l_{U}\rightarrow 0.$ Then, either $%
u^{U}\rightarrow 0$ or $l_{U}\rightarrow 0$. The first case is typical of
the BSW effect near the event horizon \cite{cqg}, while the second one is
the property of the bifurcation surface. From another hand, for a usual
particle, $\alpha _{2}\neq 0$ and $\beta _{2}\neq 0$ notwithstanding the
fact that $l_{U}\rightarrow 0$. This is because $u^{U}\rightarrow \infty $
in such a way that the product $u^{U}l_{U}$ remains finite nonzero
(actually, this is just the definition of a usual particle). The fact that $%
u^{U}\rightarrow \infty $ is consistent with the normalization \thinspace $%
u^{\mu }u_{\mu }=-1,$ it simply implies that $u_{V}\rightarrow 0.$
Geometrically, this means that a particle moves along a leg of the horizon 
\cite{cqg}. In other words, we require that for a usual particle $\beta \neq
0$ and, as a consequences, obtain the aforementioned properties of its
velocity.

It is also worth stressing that the fact that our tetrad becomes singular
near the surface of interest causes no difficulties. Indeed, we already
pointed out that the results of calculations do not depend on the choice of
the tetrad which can be subjected to rotation. Apart from this, the metric
itself remains regular near the bifurcation surface. It is also worth
reminding that for the "standard" BSW effect near the event horizon far from
the bifurcation surface the components of the four-velocity also singular in
the sense that $u^{U}\rightarrow 0$, $u^{V}\rightarrow \infty $ (or vice
versa) in the regular Kruskal-like coordinates but this does not lead to any
difficulties - instead, this is one of manifestation of the BSW effect (see
the end of Sec. 3 in Ref. \cite{cqg} for details).

\section{Exactly solvable example: Reissner-Nordstr\"{o}m metric}

To illustrate general features discussed above, it is instructive to
consider a concrete example. To test the approach, I choose the metric for
which the results were already found by quite different methods. I choose
the part of the Reisnner-Nordstr\"{o}m metric near the inner horizon and
will show that the result agrees with that obtained in \cite{in}. I express
the vectors $l^{\mu }$, $N^{\mu }$ in terms of the coordinates and, using
the equations of motion, trace how the general results of the previous
section are reproduced.

The detailed analysis of trajectories passing through in the immediate
vicinity of the bifurcation point was carried out in Sections III B and III
C of Ref. \cite{in}, so I will not repeat here the corresponding results.
Instead, I will concentrate on the properties of the null tetrad in this
vicinity and those of the particle's four-velocity which were not discussed
in \cite{in}.

The metric can be written in the form%
\begin{equation}
ds^{2}=-fdt^{2}+\frac{dr^{2}}{f}+r^{2}(d\theta ^{^{2}}+d\phi ^{2}\sin
^{2}\theta )  \label{rn}
\end{equation}%
where 
\begin{equation}
f=\left( 1-\frac{r_{+}}{r}\right) \left( 1-\frac{r_{-}}{r}\right) \text{,}
\end{equation}%
$r_{+}$ has the meaning of the event horizon radius, $r_{-}$ corresponds to
the inner horizon. We are interested in the inner region $r_{-}<r<r_{+}$. It
is convenient to rewrite the metric there as%
\begin{equation}
ds^{2}=-\frac{dT^{2}}{g(T)}+g(T)dy^{2}+T^{2}(d\theta ^{^{2}}+d\phi ^{2}\sin
^{2}\theta )\text{, }  \label{rni}
\end{equation}%
$r=-T$ is a timelike coordinate and $t\equiv y$ is spacelike, $g=-f>0$. Now,
we can choose the lightlike basis vectors as follows:%
\begin{equation}
l^{\mu }=(g,-1,0,0),l_{\mu }=(-1,-g,0,0),
\end{equation}%
\begin{equation}
N^{\mu }=\frac{1}{2g}(g,1,0,0)\text{, }N_{\mu }=\frac{1}{2g}(-1,g,0,0),
\end{equation}%
where we use the coordinates ($T,y,\theta ,\phi $). It follows from the
equations of motion that%
\begin{equation}
u^{y}=\dot{y}=\frac{X}{mg}\text{,}  \label{uy}
\end{equation}%
\begin{equation}
u^{T}=\dot{T}=\sqrt{g+\frac{X^{2}}{m^{2}}}\text{,}  \label{uT}
\end{equation}%
where dot denotes differentiation with respect to the proper time,%
\begin{equation}
X=P-\frac{qQ}{T}\text{,}
\end{equation}%
$q$ is the particle's charge, $Q$ is the charge of a black hole, $P\,\ $has
the meaning of the conserved momentum in $y$direction$.$

The proper time%
\begin{equation}
\tau =m\int_{r}^{r_{1}}\frac{dr}{\sqrt{m^{2}g+X^{2}}}\text{.}  \label{tau}
\end{equation}

We will consider two cases - $X_{-}=X(r_{-})\neq 0$ and $X_{-}=0$. We will
see below that this is completely equivalent to the choice of usual ($\beta
\neq 0$) and critical ($\beta \rightarrow 0$) particles, respectively. For
what follows, we need explicit asymptotic behavior of the four-velocity near
the horizon.

For a usual particle, one can easily obtain from (\ref{uy}), (\ref{uT}) that
in the coordinates ($T,y,\theta ,\phi $)

\begin{equation}
u^{T}\approx \frac{X(r_{-})}{m}\text{, }u^{y}=\frac{X(r_{-})}{2m\kappa
_{-}(r-r_{-})}.  \label{us}
\end{equation}%
Here, $\kappa _{-}=\frac{1}{2}\left( \frac{dg}{dr}\right) _{\mid r=r_{-}}$
has a meaning of the surface gravity of the inner horizon, it is chosen for
definiteness that $X_{-}>0.$

For the critical particle,

\begin{equation}
u^{T}\approx \sqrt{\frac{\kappa _{-}}{2}(r-r_{-})}\text{,}  \label{Tcr}
\end{equation}%
\begin{equation}
u^{y}=\frac{X}{mg}\approx \frac{X_{1}}{2m\kappa _{-}}=\left( u^{y}\right)
_{-}  \label{ycr}
\end{equation}%
where $X\approx X_{1}(r-r_{-})$ near the inner horizon, $X_{1}$ is some
constant.

The coordinates (\ref{rni}) become degenerate near the horizon since $%
g\rightarrow 0$ there. To remedy this shortcoming, one can introduce the
Kruskal-like coordinates in the standard manner. This can be done as follows:%
\begin{equation}
U=\exp [-\kappa _{-}(t-r^{\ast })]\text{,}  \label{uk}
\end{equation}%
\begin{equation}
V=\exp [\kappa _{-}(t+r^{\ast })]\text{,}  \label{v}
\end{equation}%
the tortoise coordinate%
\begin{equation}
r^{\ast }=\int \frac{dr}{g}\text{.}  \label{r}
\end{equation}%
When $r\rightarrow r_{-}$, \ the tortoise coordinate diverges, 
\begin{equation}
r^{\ast }\approx \frac{1}{2\kappa _{-}}\ln (r-r_{-})+r_{0}^{\ast }
\label{tort}
\end{equation}%
where $r_{0}^{\ast }$ is a constant. It follows from (\ref{uk}), (\ref{v})
that%
\begin{equation}
\frac{U}{V}=\exp (-2\kappa _{-}t)\text{, }UV=\exp (2\kappa _{-}r^{\ast
})=(r-r_{-})\chi (r)\text{,}  \label{uv}
\end{equation}
Near the horizon the function $\chi (r)$ is finite (its exact form is
irrelevant for our purposes), one can choose $\chi (r_{-})=1$.

Then, the metric acquires the form (\ref{f}) with%
\begin{equation}
F=g\kappa _{-}^{-2}\exp (-2\kappa _{-}r^{\ast })=\frac{g}{\kappa _{-}^{2}UV}%
\text{.}  \label{F}
\end{equation}

Now, in coordinates ($U,V,\theta ,\phi $) our basis null vectors read%
\begin{equation}
l_{\mu }=\kappa _{-}F(V,0,0,0)\text{,}  \label{lu}
\end{equation}%
\begin{equation}
N_{\mu }=\frac{1}{2\kappa _{-}}(0,\frac{1}{V},0,0)\text{.}  \label{n}
\end{equation}

The vector $l_{\mu }$ is tangent to the horizon $U=0$. It does vanish in the
bifurcation point where both $U$ and $V$ vanish. It is instructive to note
that the vector $l_{\mu }$ can be also presented in another form. It follows
from (\ref{uk}), (\ref{v}) that $\frac{\partial r}{\partial U}=\frac{\kappa
_{-}V}{2}F$, $\frac{\partial r}{\partial V}=\frac{\kappa _{-}U}{2}F$. On the
horizon $U=0$ we have that $\frac{\partial r}{\partial V}=0$. As a result,
we can write $\left( l_{\mu }\right) _{U=0}=2\left( \frac{\partial r}{%
\partial x^{\mu }}\right) _{U=0}$ Near the bifurcation point, also $\frac{%
\partial r}{\partial U}\rightarrow 0$ in agreement with discussion in the
previous Section, so $l_{\nu }\rightarrow 0$ as well.

Now, in the Kruskal coordinates, we obtain from (\ref{us}) that for a usual
particle,%
\begin{equation}
u^{U}\approx -\frac{A}{V},  \label{uuu}
\end{equation}%
where%
\begin{equation}
A=\frac{2}{F(r_{-})\kappa _{-}}\frac{X(r_{-})}{m}\text{,}
\end{equation}%
and 
\begin{equation}
u^{V}\approx \frac{2V}{F(r_{-})_{-}A}\text{.}
\end{equation}

It means for that for small $V$ a particle moves with almost vanishing $%
u^{V} $, i.e. almost along the leg $V=0$. \ Using (\ref{lu}) and (\ref{uuu}%
), one can calculate the horizon value%
\begin{equation}
\beta =-u^{\mu }l_{\mu }\approx 2\frac{X(r_{-})}{m}\neq 0
\end{equation}%
which corresponds exactly to its counterpart derived near the event horizon
(see discussion between eq.s 23 and 24 in Ref. \cite{cqg}).

For the critical particle, one finds from (\ref{uy}), (\ref{uT}) that $y$
remains finite, \ so (\ref{uv}) entails that $U\sim V\sim r-r_{-}$. Then,
one can find easily from (\ref{Tcr}), (\ref{ycr}) that $u^{U}$ and $u^{V}$
remain finite. As near the bifurcation point 
\begin{equation}
U=V=0\text{.}
\end{equation}%
the vector $l_{\mu }\rightarrow 0$ according to (\ref{lu}), we obtain that $%
\beta \rightarrow 0$. For the radial motion in the Reissner-Nordstr\"{o}m
metric $\beta =\alpha $ according to (\ref{uu}) with $s_{\mu }=0$, so $%
\alpha \rightarrow 0$ as well.

As a result, we see that the ratio $\frac{\beta _{2}}{\alpha _{1}}$ in (\ref%
{g}) diverges, so the BSW effect takes place. It is worth stressing that if
we rescale (\ref{lu}), (\ref{n}) (say, due to another reparametrization), it
does not affect the final result for the gamma factor (\ref{g})
corresponding to the energy in the centre of mass frame. Indeed, if $l_{\mu
}\rightarrow \lambda l_{\mu }$, we must change $N^{\mu }\rightarrow \lambda
^{-1}N_{\mu }$ to preserve normalization $l^{\mu }N_{\mu }=-1$. Then, $\beta
_{i}\rightarrow \lambda \beta _{i}$, $\alpha _{i}\rightarrow \lambda \alpha
_{i}$ according to (\ref{ba}), so the ratio $\frac{\beta _{2}}{\alpha _{1}}$
remains unaffected.

It is convenient to summarize the situation with the help of Table 1. Here,
for definiteness, it is implied that a usual particle moves towards the
horizon $U=0$. We compare the situation near the bifurcation surface and
near the event horizon far from it (see Sec. 3.2 of \cite{cqg}).

\begin{tabular}{|l|l|l|l|l|}
\hline
&  & usual particle & critical particle \allowbreak & $l^{\mu }$ \\ \hline
& Behavior of $\beta =-u^{\mu }l_{\mu }$ & $\neq 0$ & $\rightarrow 0$ &  \\ 
\hline
1 & Event horizon & $u^{U}$ finite & $u^{U}\rightarrow 0$ & $\neq 0$ \\ 
\hline
2 & Vicinity of bifurcation surface & $u^{U}\rightarrow \infty $ & $u^{U}$
finite\allowbreak\  & $=0$ \\ \hline
\end{tabular}

Table 1. Behavior of different factors in the expression for the coefficient 
$\beta $.

Thus in case 1 the critical particle moves almost along the horizon $U=0$
while a usual one crosses it arbitrarily. In case 2 the critical particle
passes through the bifurcation point (the near-critical passes through very
closely to it) while a usual one moves almost along the horizon $V=0.$

We see that quite different mechanisms act to ensure in each case that one
of colliding particle have $\beta \rightarrow 0$ (critical) and the other
one have $\beta \neq 0$ (usual)$.$

\section{de Sitter metric}

Another venue where the results of the present paper can be used is the BSW
effect near the cosmological horizon. It is clear from our approach that the
proof concerning the role of the bifurcation point applies to such horizons
as well. Let us consider the de Sitter spacetime as one of simplest
examples. Let the metric have the form (\ref{n}) with%
\begin{equation}
f=1-\frac{\Lambda }{3}r^{2}\text{, }\Lambda >0\text{.}
\end{equation}

In the T-region \ the metric takes the form (\ref{rni}) with%
\begin{equation}
g=\frac{\Lambda }{3}T^{2}-1=g=\frac{\Lambda }{3}(r^{2}-r_{h}^{2})\text{, }
\end{equation}%
where $r_{h}=\frac{1}{\sqrt{\Lambda }}$ corresponds to the horizon.
Equations of motion in the T-region are given by (\ref{uy}), (\ref{uT}).

Now, although the concrete formulas for the metric change, division to the
critical ($X=0$) and usual ($X\neq 0$) particles is completely similar to
that described in the previous Section. As now $X$ is the integral of
motion, $X=X(r_{h})=const$, so a particle if critical if $X=0$ and is usual
if $X\neq 0$. The quantity $X$ has the meaning of the momentum.

\subsection{Critical particle, $X=0$}

Then, it follows form (\ref{uy}) that $y=y_{0}=const$. It is seen from (\ref%
{uk}), (\ref{v}) that this trajectory passes through the bifurcation point $%
U=V=0$. The integral in (\ref{uT}) takes a very simple form:

\begin{equation}
\tau =\int_{r}^{r_{1}}\frac{dr}{\sqrt{g}}=-\sqrt{\frac{3}{\Lambda }}\ln (r+%
\sqrt{r^{2}-r_{h}^{2}})+const\text{, }
\end{equation}%
whence it is clear that $\tau -\tau _{h}\sim \sqrt{r-r_{h}}$ near the
horizon similarly to eq. (15) of \cite{in}, where $\tau _{h}$ is the value
of $\tau $ when the horizon is crossed.

\subsection{Usual particle, $X\neq 0$}

The integrals can be calculated exactly in terms of elementary functions but
the corresponding expressions are rather cumbersome and are not listed here.
It follows from (\ref{uy}) that for such a particle $y\sim \ln (r-r_{h})$
near the horizon. Therefore, in the point of collision where coordinates of
both particles should coincide, the absolute value of $y_{0}$ should also
have the order $\left\vert \ln (r-r_{h})\right\vert $ (cf. \cite{in}).

Let us consider collision of two particles, where particle 1 is critical ($%
X_{1}=0$) but particle 2 is usual ($X_{2}\neq 0$). Then, it follows from
eqs. (\ref{e}) and (\ref{gamma}) and equators of motion (\ref{uy}), (\ref{uT}%
) that%
\begin{equation}
E_{c.m.}^{2}=m_{1}^{2}+m_{2}^{2}+\frac{m_{1}Z_{2}}{\sqrt{g}}\text{, }Z_{2}=%
\sqrt{X_{2}^{2}+m_{2}g}
\end{equation}%
Thus in the horizon limit $g\rightarrow 0$ the energy in the centre of mass
frame does diverge. We would like to stress one more time that for getting
growing energies it is necessary that one of particles be critical. In turn,
it means that it passes through the bifurcation point as is explained above.
The closer the point of collision to the bifurcation point for a fixed $%
X_{2} $, the larger $E_{c.m.}^{2}$.

It is worth noting that collision of particles near the inner horizon of the
de Sitter - Reissner-Nordstr\"{o}m metric (with the electric charge $Q\neq 0$%
) was considered in \cite{srn}. In doing so, it was essential that a
critical particle was also electrically charged. However, the role of the
bifurcation surface was not revealed there. If this is done, it follows from
our consideration that the BSW effect occurs even if $Q=0$ and particles are
neutral.

\section{Conclusion}

Thus we have managed to generalize the BSW effect for the vicinity of an
arbitrary bifurcation surface not using the details of the metric. Using the
Reissner-Nordstr\"{o}m metric as an example, we checked that this approach
and that based on equations of motion agree with each other. Meanwhile, now
the knowledge of these equations is not required and was used for comparison
of both approaches only. The corresponding conclusions apply to the inner
region of nonextremal black holes having an inner horizon. What is
especially important, they are also valid for cosmological horizons. Here,
we also compared the results obtained in a general approach with that based
on equations of motion. This confirmed that the BSW effect near the
bifurcation point does exist.

At present, it is far from being clear whether and how this effect can
influence cosmological evolution but, in any case, the present results can
serve as motivation to pose such a question.

To summarize, we showed that a generic bifurcation surface is an accelerator
of particles in a strong gravitational field.

\end{document}